\documentstyle[preprint,revtex]{aps}

\begin{document}
\thispagestyle{empty}


\begin{center}
\hfill{IP-ASTP-02-94}\\
\hfill{April, 1995}

\vspace{1 cm}

\begin{title}
Eigenstates of the Vorticity Operator:
The Creation and Annihilation of Vortex States in Two Dimensions
\end{title}
\vspace{0.5 cm}

\author{A.~D.~Speliotopoulos}
\vspace{0.10 cm}
\begin{instit}
Institute of Physics, Academia Sinica\\ Taipei, Taiwan 115, R.O.C.
\end{instit}
\end{center}

\begin{abstract}

\begin{center}
{\bf Abstract}
\end{center}

\noindent A quantum mechanical description of vortices in
 two dimensional superfluid $^4$He flims is presented. Single vortex
 creation and annihilation operators are defined and wavefunctions for
 these states are explicitly constructed. A hamitonian for these
 states is then proposed.

\end{abstract}
\vfill
\vfill
\newpage

\noindent{\bf \S 1. Introduction}

As is well known by now, the phase transition of the two dimensional
superfluid $^4$He film is quite well described by the Kosterlitz-Thouless
(KT) phase transition $\cite{1}$-$\cite{4}$ which makes
explicit use of the presence of a gas of point vortices in the
superfluid. As the dynamical properties of these vortices must be
known before the critical properties of the phase transition can be
calculated, an analogy is made between the superfluid and the
classical ideal liquid. Kirchoff's equation of vortex motion can then
used to define a grand canonical ensemble for the system and the
critical indices can be calculated. Superfluidity, however, is
believed to be a quantum mechanical phenomenon while KT phase transition
relies on {\it classical\/} fluid dynamics $\cite{5}$ to construct
a classical grand canonical ensemble. The only place quantum mechanics is
used is in the requirement that the ``charge'' or vorticity of these
vortices must be integral multiples of $h/m$. It is
therefore quite natural to wonder if a quantum mechanical description of
these vortices is possible.

The purpose of this paper, therefore, is to develope a quantum
mechanical description of vortices in superfluid $^4$He films. There
are two approaches that one can take. The most straightforward, naive
approach would be to continue treating the vortices as point
particles in and of themselves and attempt to ``quantize''
their classical equations of motion directly. When one tries to do
so, however, one is immediately presented with seemingly
insurmountable problems caused, ironically enough, by the very
simplicity of Kirchoff's equations themselves. If we label the
coordinate of a vortex by $(x,y)$, then from Kirchoff's equations
the momentum conjugated to $x$ is the coordinate
$y$. As Onsager $\cite{7}$ first noted, naive canonical quantization
would then seem to give the non-sensical result that the two coordinates
do not commute: $[x,y]\ne0$. Moreover, because the classical
equations of motion depend only on the velocity of the vortices,
their lagrangian is {\it linear\/} in their velocities. Consequently,
their classical hamiltonian does not contain a kinetic piece, but is
instead all potential. One does not have a ``free''
hamiltonian which can be perturbed about and must instead
deal immediately with the fully interacting theory. Nevertheless,
this approach has been attempted by McCauley $\cite{8}$ who
implimented the canonical quantization condition by everywhere
replacing, by hand, the coordinates $x$ and $y$ of the vortices with
the raising and lowering operators of the SHO. Although certain
results, such as the finite core size of  the vortex, can be
obtained in this manner, this replacement is ad~hoc and only a system
containing at most two vortices with fixed circulations can easily be
treated by it. The total vorticity is necessarily fixed in this
approach. In addition, it is not altogether certain that the {\it two
dimensional\/} vortex gas can be consistently quantized in this
manner, as has been argued in $\cite{9}$, $\cite{10}$.

The approach we shall follow in this paper is based on the
observation that we already know which states of the fluid are
important and necessary to describe the superfluid phase transition.
Namely, they are states of definite vorticity. With the states of the
fluid known, we then need only find the relevant physical operators
for these states. By considering the vorticity as a
good quantum number for the system we can define a vorticity operator
$(h/m){\cal Q}$ for which these states are eigenstates of. The
eigenvalues of ${\cal Q}$ shall then simply be the vorticity of the
system in units of $h/m$. Using complex coordinates, it is
straightforward to develope a {\it heuristic\/} derivation of these
states and, through them, a coordinate representation of ${\cal Q}$.
With this in hand, we shall, as in the case of the simple harmonic
(SHO), be able to define ladder operators, creation and annihilation
operators, $c$ and $c^\dagger$, whose operation on eigenstates of
${\cal Q}$ takes one from one state of definite vorticity to another.
A hamiltonian $H_c \sim c^\dagger c$ can then be defined, which,
importantly, {\it commutes\/} with ${\cal Q}$. States of definite
vorticity in the fluid may then be eigenstates of both $H_c$ and
${\cal Q}$. Moreover, the algebra of operators {\it closes}, and no
other operators need to be introduced. As the wavefunctions and the
relevant operators are known, the quantization of the single vortex
system is then complete.

This approach is the very reverse of the
proceedure one usually follows when quantizing a classical theory. One
usually starts with a hamiltonian and construct from it the
wavefunctions and Hilbert space. In our case, we know what the relevant
states of the system are, and from these states we shall construct
the hamiltonian and the other relevant physical operators of the
theory. This approach has obvious draw backs, of course, not the
least of which will be a hamiltonian which is significantly different
from the one one usually encounters. We can only be certain that this
hamiltonian is the relevant one by showing that it, and its
eigenvalues, has many of the same properties as the classical
hamiltonian. It is also only in this way that the unknown constants
in the theory can be fixed.

The rest of this paper is organized in the following manner. {\bf
\S 2} is devoted to reviewing certain properties of classical
vortices, in particular the difficulties inherent in the simplistic
canonical quantization of the vortex gas. In $\bf \S 3$ we shall
define and construct the algebra of operators which are relevant to
the description of a single point vortex in the superfluid. Following
the analysis of the SHO, the single vortex Hilbert space shall be
constructed. This construction is purely algebraic, however, and we
shall delay showing that a faithful representation of the algebra can
be chosen until $\bf \S 4$ where the single vortex wavefunctions will
be constructed and interpreted. Concluding remarks can be found in
$\bf \S 7$.

A brief comment on terminology. Because vortices in quantum liquids
have ``quantized'' circulation, one often refers to these vortices as
``quantum vortices''. This is to differentiate these vortices from
vortices found in classical fluids whose circulations need not be
integer valued. This terminology may, however, be confusing and
misleading in the current context since in no way are the motion of
vortices quantized. Because of its usefulness, however, and its
prevelance in the literature, we also shall use it. Nevertheless, we
caution the reader against any possible misunderstanding which may
result.

\noindent{\bf \S 2. Background}

We begin with a brief review of the standard theory and
application of quantum vortices to superfluidity. This review is
necessarily cursury and the reader is refered to $\cite{11}$
for further details. Although we are working with vortices in
two dimensional films, we shall begin by considering vortex lines in
three dimensions. This is partly because the
traditional argument for the presence of point vortices in two
dimensional films follows in direct analogy to the three dimensional case.
Mainly, however, it is because certain problems and inconsistencies
araises in this argument when one actually tries to impliment it in two
dimensions $\cite{12}$-$\cite{14}$. We shall delay discussion of
such problems until later in the section, however, and
shall gloss over for now  any problems which may araise.

We start with the standard definition of what is meant by a quantum
vortex in three dimensional $^4$He superfluids. Let $\psi$ be the
microscopic bosonic field for $^4$He and consider its
thermodynamic average $\langle\psi\rangle$. In the normal fluid
phase this average vanishes due to the $U(1)$ gauge invariance of the
hamiltonian for the theory. In the superfluid phase, on the other
hand, $\langle\psi\rangle\ne 0$, denoting the breaking of the
$U(1)$ gauge symmetry. A current density for $\langle\psi\rangle$
can then be constructed as
\begin{equation}
\vec j = {\hbar\over 2mi}
	\left(
		\langle\psi^\dagger\rangle\vec\nabla\langle\psi\rangle
		-
		\vec\nabla\langle\psi^\dagger\rangle\langle\psi\rangle
	\right)\>,
\label{e2.1}
\end{equation}
where $m$ is the mass of the boson. If we then polar decompose
$\langle\psi\rangle = \rho^{1/2}_s\exp(i\alpha)$,
\begin{equation}
\vec j = {\hbar \rho_s\over m}\vec\nabla\alpha\>.
\label{e2.2}
\end{equation}
Then, because the microscopic field is a boson,
\begin{equation}
\oint_\gamma {\vec j\over \rho}\cdot d\vec l = 2\pi q{\hbar\over m}\>,
\label{e2.3}
\end{equation}
where $\gamma$ is any closed path in the fluid and $q$ is an {\it
integer}. From analogy with fluid dynamics, eq.~$(\ref{e2.3})$ is called
the {\it circulation\/} of a vortex. Because $\vec j$ is
proportional to a total derivative, the circulation
is independent of the shape $\gamma$. Since $q$ is an {\it integer}, at
this point one usually says that there is a {\it quantum\/} vortex
with circulation $q$ somewhere within $\gamma$ when $q\ne0$. As,
however, $\gamma$ is arbitrary, one can conceiveably choose a path
which will encompass more than vortex within the loop. It is therefore
always tacitly assumed that this path $\gamma$ will encircle one and
only one vortex in the fluid.

To describe the dynamical motion of these vortices, one usually
appeals to classical fluid dynamics. Because the superfluid component
of the fluid flow in three dimensions is non-dissipative in nature,
it behaves much like a classical ideal fluid. As ideal point
vortices may also be present in the ideal fluid, one may then directly
carry over and apply to quantum vortices in the superfluid the classical
equations of vortex motion. Based on this argument, any
motion of quantum vortices in the superfluid can be analyzed
using {\it classical\/} fluid dynamics.

Let us now restrict ourselves to the case of two dimensions. {\it
Formally\/} the above observations still hold, although we shall
comment on certain problems with this argument latter on. Instead
of having to work with vortex lines, we are now left with only point
vortices, a very welcome simplification. In
particular, by labeling the locations of the cores of the vortices by
$\vec x^\alpha$, $\alpha = 1, \dots N$, the number of vortices in the
fluid, we can treat the vortices as though they were point particles in and
of themselves instead of being manifistations of a
current flow in the film. Then by working in complex coordinants
$z^\alpha = x^\alpha + iy^\alpha$, and ${\bar z}^\alpha = x^\alpha -
iy^\alpha$, we appeal once again to fluid dynamics $\cite{5}$ and find that
their lagrangian is
\begin{equation}
L = i\pi \hbar\rho_0\sum^N_{\alpha=1} q_\alpha
		\left(
		{dz^\alpha\over dt}{\bar z}^\alpha
		-
		{d{\bar z}^\alpha\over dt}z^\alpha
		\right)
	- K\>,
\label{e2.5}
\end{equation}
where
\begin{equation}
K = -e_{kt} \sum^N_{q_\alpha\ne q_\beta} \log
	\left\vert z^\alpha-z^\beta\over a_s\right\vert^2,
\label{e2.6}
\end{equation}
and $q_\alpha$ is the circulation of a vortex located at position
$z^\alpha$ in the fluid. $a_s$ is some length scale for the system
which is often identified as the size of the vortex core and $\rho_0$ is
the average {\it total \/} density of the film while $e_{kt}$ is
the KT energy scale: $e_{kt}= \rho_0
(2\pi\hbar)^2/(4m)$. The infinite self-energy terms of the vortices have
not been included in eq.~$(\ref{e2.6})$. (One does not have to use
fluid dynamics to obtain eq.~$(\ref{e2.6})$. It can also be derived
using the lagrangian for the microscopic field $\psi${} $\cite{15}$,
$\cite{16}$.)

Now that we have a lagrangian for a ``gas'' of vortices in a two
dimensional fluid, we can try to ``quantize'' this theory using
naive canonical quantization. To do so, one takes as the generalized
coordinate ${\hbox{\bf q}}^\alpha = z^\alpha$. Its canonical momentum
is then ${\hbox{\bf p}}_\alpha = 2\pi i\hbar\rho_0q_\alpha z^\alpha$
and their Possion bracket is
\begin{equation}
\{z^\alpha, {\bar z}^\beta\} =
-i\delta_{\alpha\beta}/(2\pi\hbar\rho_0 q_\alpha)\>,
\label{e2.7}
\end{equation}
which makes perfectly good sense at a classical level. Next, because
the lagrangian is linear in the velocities, we find that the
classical hamiltonian $H_{cl}$ for the vortices is
\begin{equation}
H_{cl} = \sum^N_{\alpha,\beta} {\hbox{\bf p}}_\alpha {d{\hbox{\bf
q}}_\alpha\over dt} - L = K\>,
\label{e2.8}
\end{equation}
and does not contain a kinetic piece but is instead all ``potential''.

If we now try to quantize this theory using naive canonical
quantization, we immediately run into problems. First, the
hamiltonian is all ``potential'' and is fully interacting. It does
not have a ``free'' part about which we can perturb. Second, because the
lagrangian is linear in the velocities, one may need to use Dirac's
theory of contraints $\cite{17}$ while quantizing the theory to project out
the unphysical states. Neither of these problems would seem to be
insurmountable, however, until one tries to impliment canonical
quantization; namely replacing the Poisson bracket by a commutator
between two operators: $[z^\alpha, {\bar z}^\beta] =
\delta_{\alpha\beta}/(2\pi\rho_0q_\alpha)$. This, however,
immediately implies that $z^\alpha$ does not commute with its complex
conjugate ${\bar z}^\alpha$. Seemingly, the spatial coordinates of the quantum
theory do not commute among themselves. Although technically this
does not seem to present any great difficulties for two vortices
$\cite{8}$, it does present certain conceptual problems and,
fundamentally, is the reason why the vortex gas has yet to be
successfully quantized. Moreover, from the work by $\cite{9}$ and
$\cite{10}$, it is very doubtful that this naive quantization of the
vortex gas can succeed.

Notice that the above arguments, expecially the form that $L$ takes,
relies heavily on the analogy made between the classical
ideal fluid and the superfluid. Actually, any and all quantum
mechanical input into the subsequent arguments ends with the
quantization condition eq.~$(\ref{e2.3})$. As we have remarked, however,
that although these arguments in support of the analogy hold quite
well in three dimensions, the situation is not nearly as clear cut in
two dimensions. This is because there are well known theorems which
state that in two dimensions, $\langle \psi\rangle=0$ at all
temperatures $\cite{12}$-$\cite{13}$. It is also known, however, that
the KT transition, which is based on the presence of
point vortices in the two dimensional superfluid, is a good model of
the superfluid transition. Consequently, we shall assume that these
vortices are present in the fluid and can be described at least
heuristically in the same way that vortices in three dimensional fluids are.

\noindent{\bf\S 3. The Single Vortex Hilbert Space}

We now begin the construction of the single vortex Hilbert space. As
this construction is somewhat long and convoluted, we will first give
an overview of how this is done. Fundamentally, what we shall be
doing is determining explicitly the configuration of the fluid which
will give rise to quantum vortices. First, by working with
complex coordinates, a specific representation of the quantization
condition eq.~$(\ref{e2.3})$ is found in terms of functions on the complex
plane. From it, the approximate form of the wavefunction describing a
single vortex in the fluid with definite vorticity is obtained
heuristically. A vorticity operator $(h/m){\cal Q}$ is then defined
which, when acting on a state of definite vorticity, returns simply
its vorticity, an integer. Since this operator may be factorized into
two pieces, we can define creation $c$  and annihilation operators
$c^\dagger$ for the single vortex states. With the addition of the
operator $c^\dagger c$, the algebra of relevant operators closes.
Treating this algebra as fundamental and given, the single vortex
Hilbert space is then constructed algebraicly in direct analogy with
the SHO.

We begin by finding the relevant operators using the quantization
condition eq.~$(\ref{e2.3})$. To do so, we recast the quantization
condition
\begin{equation}
\frac{h}{m} q = \int_\gamma \frac{j_z}{\rho} dz\>,
\label{e3.1}
\end{equation}
as a contour integral in the complex plane where
\begin{equation}
j_z = \frac{\hbar}{2mi}\left(
	\langle \psi^\dagger\rangle
	\frac{\partial\langle\psi\rangle}{\partial z}
	-
	\frac{\partial\langle\psi^\dagger\rangle}{\partial z}
	\langle \psi\rangle \right)\>.
\label{e3.2}
\end{equation}
In doing so, we are saying that $j_z/\rho$ is a meromorphic function
while $j_{\bar z}/\rho$ gives no additional information and is
redundent. Written in this form, we would expect heuristically that
the wavefunction which describes a single vortex in the fluid with
vorticity $q$ to have the form
\begin{equation}
\phi \sim \rho^{1/2}\left(\frac{z}{\bar z}\right)^q\>,
\label{e3.3}
\end{equation}
where for convenience we now represent the wavefunctions of the fluid
by $\phi$ instead of $\langle \psi\rangle$.

We then define the vorticity operator $(h/m){\cal Q}$ as being that
operator for which any wavefunction satisfying eq.~$(\ref{e3.1})$ is
an eigenstate of ${\cal Q}$. Namely, ${\cal Q}$ operating on a $\phi$
of the form given in eq.~$(\ref{e3.3})$ gives: ${\cal Q}\phi=q\phi$.
By definition, an eigenstate of this
operator satisfies the quantization condition eq.~$(\ref{e3.1})$ and visa
versa. Using ${\cal Q}$ we may now directly impliment the
quantization condition in terms of an operator acting on states of a
Hilbert space.

With the form of $\phi$ given in eq.~$(\ref{e3.3})$, we can choose a
specific representation of ${\cal Q}$ in terms of differential operators
\begin{equation}
{\cal Q} = {1\over2}\left(z{\partial \>\>\>\over\partial z} -
{\bar z}{\partial \>\>\>\over\partial {\bar z}} \right)\>,
\label{e3.8}
\end{equation}
as long as $\rho$ is a function of $\vert z\vert$ only. From this we
would naively expect ${\cal Q}$ to be an hermitian operator, as
required. Notice also that it is proportional to the generator of
rotations in two  dimensions. We next immediately see that it can be
factorized
\begin{equation}
{\cal Q} = \left[{1\over\sqrt2}z{\partial \>\>\>\over\partial {\bar z}},
	{1\over\sqrt2}{\bar z}{\partial \>\>\>\over\partial z}\right]\>,
\label{e3.9}
\end{equation}
into a commutator of two other differential operators.
With this observation it is then straightforward to construct
the operators
\begin{equation}
c \equiv {1\over \sqrt2} \left({{\bar z}\over
z}+{\bar z}{\partial \>\>\>\over\partial z}\right)
\qquad,\qquad
c^\dagger \equiv {1\over \sqrt2} \left({z\over
{\bar z}}-z{\partial \>\>\>\over\partial {\bar z}}\right)\>,
\label{e3.10}
\end{equation}
which, as we shall see, will serve as creation and annihilation
operators for the vortex states. From their form
we would {\it expect\/} $c$ and $c^\dagger$ to be adjoints of one
another. Unfortunately, the situation is not nearly as clear cut, but
for the sake of clarity we shall delay addressing this issue until
$\bf \S 4$. For now, we shall simply assume that they are
truly adjoints of each other.

Defining $H_c = \epsilon c^\dagger c$, we find that $c, c^\dagger,
H_c$ and ${\cal Q}$ have the following commutation relations:
\begin{eqnarray}
[c, c^\dagger] = I + {\cal Q}\qquad&,&\qquad [H_c, {\cal Q}]=0\>,
\nonumber \\
{[c, {\cal Q}]} = c\qquad&,&\qquad [H_c, c]=-\epsilon(I+{\cal Q})c\>,
\nonumber \\
{[c^\dagger, {\cal Q}]} = -c^\dagger \qquad&,&\qquad [H_c, c^\dagger] =
\epsilon c^\dagger (I+{\cal Q})\>.
\label{e3.11}
\end{eqnarray}
where $\epsilon$ is a constant having units of energy and shall be
identified later. For reasons that will be made clearer in $\bf \S
4$, we identify $H_c$ as the single vortex hamiltonian.
For now, we simply note that ${\cal Q}$ commutes with $H_c$ and,
consequently, is a conserved charge of the system. Notice also that
its commutation relations with $c$ and $c^\dagger$ are precisely what
one would expect for a charge operator.

{}From eq.~$(\ref{e3.11})$ we see that the algebra ${\cal A} = \{I, c,
c^\dagger, {\cal Q}, H_c\}$ {\it closes}. No other operators need to be
introduced. We shall therefore take $\cal A$ as the basic set of
physically relevant operators which are needed to describe a single
vortex in a two dimensional superfluid. In addition, now that we have
found the relevant operators for the single vortex system, we shall
turn things around. Namely, we shall consider the abstract algebra
eq.~$(\ref{e3.11})$ as {\it given\/} and {\it fundamental\/} with
$c^\dagger$ {\it defined\/} as the adjoint of $c$ and ${\cal Q}$ {\it
defined\/} to be an hermitian operator. The explicit forms
of these operators given in eqs.~$(\ref{e3.8})$ and $(\ref{e3.10})$ are to be
considered as specific representations of this algebra in terms of
linear partial differential operators.

We do so because there are certain subtleties
involved in finding faithful representation of eq.~$(\ref{e3.11})$. Namely,
apprearences notwithstanding, with the representation given in
eq.~$(\ref{e3.10})$ $c$ and $c^\dagger$ are not truly adjoints of one
another. Although this problem can be resolved, doing so will take us
somewhat afield. We shall therefore delay showing that a faithful
representation of eq.~$(\ref{e3.11})$ can be found until the next section
and shall instead proceed with an algebraic construction of the
single vortex Hilbert space based on this algebra.

Because ${\cal Q}$ and $H_c$ commute, one can find simultaneous eigenstates
of both. We therefore provisionally define our single vortex
Hilbert space ${\cal H}_s$ as being those states spaned by
eigenstates of ${\cal Q}$ and $H_c$.  Further constraints shall be put on
${\cal H}_s$ as needed. To charactorize the states of ${\cal H}_s$, we
proceed in analogy with the SHO. Let $\vert\phi\rangle\in{\cal H}_s$
be an eigenstate of both ${\cal Q}$ and $H_c$ with eigenvalues $\lambda_q$
and $\lambda_c$ respectively. Then it is straightforward to show that
for any positive integer $n$,
\begin{eqnarray}
{\cal Q} c^n\vert\phi\rangle &=&(\lambda_q-n)c^n\vert\phi\rangle\>,
\nonumber \\
{\cal Q}{c^\dagger}^n\vert\phi\rangle&=&
	(\lambda_q+n){c^\dagger}^n\vert\phi\rangle\>,
\nonumber \\
H_cc^n\vert\phi\rangle &=& \left(\lambda_c + {\epsilon\over
2}n(n-1)-\epsilon n\lambda_q\right)c^n\vert\phi\rangle\>,
\nonumber \\
H_c{c^\dagger}^n\vert\phi\rangle &=& \left(\lambda_c + {\epsilon\over
2}n(n+1)+\epsilon n\lambda_q\right){c^\dagger}^n\vert\phi\rangle\>.
\label{e3.12}
\end{eqnarray}
{}From this we note the following:
\vskip20truept
\noindent 1. If $\vert\phi\rangle$ is an eigenstate of ${\cal Q}$ and $H_c$,
then $c^n\vert\phi\rangle$ and ${c^\dagger}^n\vert\phi\rangle$ are
also eigenstates of ${\cal Q}$ and $H_c$.
\vskip20truept
\noindent 2. The operator $c$ (${c^\dagger}$) operating on
$\vert\phi\rangle$ will produce a state whose ${\cal Q}$-eigenvalue has been
decreased (increased) by $1$. Since eigenstates of ${\cal Q}$ are states
with definite vorticity, we can physically interpret $c$ as the
annihilation operator for an unit of {\it positive\/} vorticity $+1$
while $c^\dagger$ is the creation operator for an unit of positive vorticity.
\vskip20truept
\noindent 3. If $\lambda_q$ is an integer for any eigenstate of ${\cal Q}$,
then all the eigenvalues of ${\cal Q}$ are integers.
\vskip20truept
\noindent 4. Like the SHO, the eigenvalues of $H_c$ must be greater then
zero. Unfortunately, this does not put as great a constraint on the
spectrum of $H_c$ as it did for the SHO since the $H_c$-eigenvalues
of both $c^n\vert\phi\rangle$ and ${c^\dagger}^n\vert\phi\rangle$
increases quadratically with $n$. No matter what the initial values
of $\lambda_c$ and $\lambda_q$ are, as long as
\begin{equation}
-(\lambda_c+1)\le \lambda_q\le\lambda_c\>,
\label{e3.13}
\end{equation}
for any one eigenstate of $H_c$ of the Hilbert space, all eigenstates
of $H_c$ will have non-negative eigenvalues. One does not
automatically obtain the quantization of the spectrum for $H_c$ as in
the case of the SHO. We shall have to appeal to physical reasoning
instead.
\vskip20truept

Suppose that we are given a film which is completely at rest in the
laboratory frame. Then there should be a state of the Hilbert space
which represents the state of the fluid in which no vortices are
present whatsoever. The fluid is completely quiescent. Let us denote
this state by $\vert0\rangle$, which we shall call the {\it ground
state\/} of the system. We {\it require}, on physical grounds, that
it be present in ${\cal H}_s$. Then ${\cal Q}\vert0\rangle=0$ and
from the above all the eigenvalues of ${\cal Q}$ for states in ${\cal
H}_s$ are quantized.

Unfortunately, determining $\lambda_q$ does not put any
bound  whatsoever on $\lambda_c$ and we shall have to use further
arguments. Since $\vert0\rangle$ represents the fluid at rest without
any vortex excitations whatsoever, it must be the state of lowest energy.
Any other state of the system containing vortices must have a higher
energy than it does. Suppose, then, that $H_c\vert0\rangle=
\lambda_0\vert0\rangle$ where $\lambda_0\ne0$. Then
$c\vert0\rangle\ne0$ and is instead a state with vorticity $-1$ from
eq.~$(\ref{e3.12})$. Moreover, we see that $H_c
c\vert0\rangle=\lambda_0\vert0\rangle$. The state with vorticity $-1$
would thus have the same energy as the state with no vorticies
whatsoever. As this is physically unreasonable, we shall require that
$H_c\vert0\rangle =0$ also. Then from the definition of $H_c$, we
find that $\big\vert c\vert0\rangle\big\vert^2=0$ and
$c\vert0\rangle=0$. $c$ annihilates the ground state and the only
relevant states in ${\cal H}_s$ are linear combinations of
${c^\dagger}^n\vert0\rangle$.

With the presence of this ground state in ${\cal H}_s$, all the eigenvalues
of ${\cal Q}$ and $H_c$ are quantized and can be enumerated by a single
quantum number $n\ge0$. The single vortex Hilbert space ${\cal H}_s$
is spanned by the states
\begin{equation}
\vert n\rangle \equiv {{c^\dagger}^n\vert0\rangle\over
\sqrt{\left(n(n+1)/2\right)!}}\>,
\label{e3.14}
\end{equation}
constructed from the ground state. From eq.~$(\ref{e3.12})$ they have
eigenvalues
\begin{equation}
{\cal Q}\vert n\rangle = n\vert n\rangle \qquad,\qquad
H_c\vert n\rangle = {\epsilon\over 2}n(n+1)\vert n\rangle\>.
\label{e3.15}
\end{equation}
The vorticity of the state $\vert n\rangle$ is therefore $nh/m$, as
expected.

We would therefore seem to be finished. Notice, however, that ${\cal
H}_s$ contains only states of positive vorticity. Due to physical
constraints on the energy of the ground state, all the negative
vorticity states were removed from ${\cal H}_s$.
This still leaves open the problem of the construction of the negative
vorticity Hilbert space, however. Fortunately, doing so
is straightforward.

Let us define the operator $P$ such that
\begin{equation}
P^2 =1 \qquad,\qquad P{\cal Q} P = -{\cal Q}\>,
\label{e3.16}
\end{equation}
which is the parity operator as can be seen explicitly
using the coordinate representation of ${\cal Q}$. We then define the
operators
\begin{equation}
d\equiv PcP\qquad,\qquad d^\dagger \equiv Pc^\dagger P\qquad,\qquad
H_d\equiv PH_cP\>,
\label{e3.17}
\end{equation}
and find that
\begin{eqnarray}
[d, d^\dagger] = I - {\cal Q} \qquad&,&\qquad [H_d, {\cal Q}] = 0\>,
\nonumber \\
{[d, {\cal Q}]}= -d \qquad&,&\qquad [H_d, d] = -\epsilon(I-{\cal Q})d\>,
\nonumber \\
{[d^\dagger, {\cal Q}]} = d^\dagger \qquad&,&\qquad [H_d, d^\dagger] =
\epsilon d^\dagger(I-{\cal Q})\>.
\label{e3.18}
\end{eqnarray}
Proceeding just as before, we find that the only relevant states are
of the form
\begin{equation}
\vert m\rangle =
{{d^\dagger}^m\vert0\rangle\over\sqrt{\left(m(m+1)/2\right)!}}\>,
\label{e3.19}
\end{equation}
where once again $\vert0\rangle$ is the ground state of the system.
Moreover, $\vert m\rangle$ is an eigenstate of ${\cal Q}$ and $H_d$
with eigenvalues
\begin{equation}
{\cal Q}\vert m\rangle = -m\vert m\rangle\qquad,\qquad H_d\vert m\rangle =
{\epsilon\over 2} m(m+1)\vert m \rangle\>.
\label{e3.20}
\end{equation}
The states $\vert m\rangle$ all have {\it negative\/}
vorticity, but with the {\it same\/} dependence of the energy
eigenvalues on $m$. We then define the Hilbert space $\bar{\cal H}_s$ as
being spaned by simultaneous eigenstates of ${\cal Q}$ and $H_d$.
Moreover, from eq.~$(\ref{e3.20})$ we can physically interpret the
operator $d$ ($d^\dagger$) as the annihilation (creation) operator
for an unit of {\it negative\/} vorticity. With the use of a parity
operator $P$ we have thus mapped the algebra ${\cal A}$ acting on the
single vortex Hilbert space ${\cal H}_s$, containing states of
positive vorticity only, into $\bar{\cal A} = \{I, d, d^\dagger,
{\cal Q}, H_d\}$ acting on $\bar{\cal H}_s$, containing only states
with negative vorticity.

Let us now return to the question of the determination of the ground
state of the Hilbert space. Remember that we {\it required\/} that
there be a state in the Hilbert space, identified as the ground
state, which is rotationally invariant. We shall now
show that this is equivalent to there being a state of the Hilbert
space which is parity invariant. Suppose that
$\vert\phi\rangle\in{\cal H}$ such that
$\vert\phi\rangle=P\vert\phi\rangle$. Moreover, suppose that
$\vert\phi\rangle$ is an eigenstate of ${\cal Q}$, so that
\begin{equation}
{\cal Q}\vert\phi\rangle = \lambda_q\vert\phi\rangle\>.
\label{e3.21}
\end{equation}
Then multiplying both sides with $P$
\begin{equation}
P{\cal Q} PP\vert\phi\rangle = \lambda_qP\vert\phi\rangle\>.
\label{e3.22}
\end{equation}
Since $P{\cal Q} P=-{\cal Q}$, we find that $\lambda_q = -\lambda_q$, and
$\lambda_q=0$. Thus, the only eigenstate of ${\cal Q}$ which is also parity
invariant has vanishing eigenvalue. Physically, our requirement that
there exists a state of the Hilbert space which has zero
${\cal Q}$-eigenvalue is simply the requirement that there exists a
state of the Hilbert space which is parity invariant. Moreover, this
state may function as the ground state for both ${\cal H}_s$ and
$\bar{\cal H}_s$.

\noindent{\bf \S 4. Single Vortex State Wavefunctions}

The construction of the Hilbert space done in the previous section,
while valid, was formal and algebraic in nature. In its construction
we {\it assumed\/} that an inner product on the Hilbert space has
already been defined and that a faithful representation of the
algebra can be chosen. These assumptions will now be justified. Then,
with the faithful representation of the algebra known, we shall
explicitly construct the single vortex wavefunctions and physically
interpret the results which were obtain abstractly in the previous
section. As the negative vorticity states can be obtained from the
positive vorticity states using the parity operator, we concern
ourselves with only positive vorticity states in this section.

Since we originally started out with a specific choice of $c$ and
$c^\dagger$ when constructing eq.~$(\ref{e3.11})$, finding a faithful
representation of this algebra would seem to be straightforward.
One would only have to find the wavefunctions by solving a first order
differential equation and then varify that the $c$ and $c^\dagger$
defined from eq.~$(\ref{e3.10})$ are truly adjoints of each other.
Unfortunately, appearances notwithstanding, they are not. If we
try using the representation of $c$ and $c^\dagger$ given in
eq.~$(\ref{e3.10})$ to construct the wavefunctions, we would obtain
wavefunctions which are singular at $\vert z\vert=0$. Due to this
singularity they will not be normalizeable. If we try to correct
for this by removing a disk of fixed radius from the origen and
restricting the domain of the wavefunctions, the wavefunctions will
become normalizeable, but we will then find that $c$ and $c^\dagger$ are no
longer adjoints of each another. The removal of this disk
generates a surface term when performing an integration by parts.
As this surface term does not vanish, in this representation $c$ and
$c^\dagger$ are not adjoints of one another and eq.~$(\ref{e3.10})$
cannot be a faithful representation of the algebra.

Curiously enough, $H_c$, defined though eq.~$(\ref{e3.10})$, and the
representation of ${\cal Q}$ in eq.~$(\ref{e3.8})$ {\it are\/} hermitian
operators even when the domain of the
wavefunctions are restricted. To show that they are hermitian
operators, two integrations by parts must be preformed,
resulting in two surface terms which cancel one another. This
suggests that the failure of eq.~$(\ref{e3.10})$ to form a faithful
representation of the algebra is not a fatal one and only slight
modifications to eq.~$(\ref{e3.10})$ only will be needed.

We begin by defining the type of wavefunctions we shall be
constructing and their inner product on the single vortex Hilbert space
${\cal H}_s$. The domain of the wavefunctions will be the complex plane
$\bf C$ with the inner product on the Hilbert space defined as
\begin{equation}
\langle \phi\vert{\cal O}\phi\rangle \equiv \int_{\bf C}
\phi^\dagger{\cal O}\phi \>d^2x\>,
\label{e4.1}
\end{equation}
for any operator $\cal O$ on the Hilbert space and $\phi\in {\cal H}_s$.
In addition, the wavefunctions are required to be $L^2$ integrable over
$\bf C$, meaning that
\begin{equation}
\int_{\bf C}\vert\phi\vert^2\> d^2x <\infty\>,
\label{e4.2}
\end{equation}
is finite for all $\phi\in\cal H$.

To find the correct representation of the algebra, we return to
eq.~$(\ref{e3.10})$ and consider its possible generalization by taking
\begin{equation}
c = {1\over \sqrt 2}\left({{\bar z}\over z}g(z, {\bar z}) +
h({\bar z}){\bar z}{\partial\>\>\over \partial z}\right)
\quad,\quad
c^\dagger = {1\over \sqrt 2}\left({z\over {\bar z}}\overline{g(z,{\bar z})} -
\overline{h({\bar z})}z{\partial\>\>\over \partial {\bar z}}\right)\>,
\label{e4.3}
\end{equation}
where $g(z,{\bar z})$ is a function of both $z$ and ${\bar z}$ while for
$c^\dagger$ to be the adjoint of $c$, $h({\bar z})$ can be a function of
${\bar z}$ only. We now require that $c$ and $c^\dagger$ be a faithful
representation of the algebra eq.~$(\ref{e3.11})$. Taking once again
eq.~$(\ref{e3.8})$
as the representation of ${\cal Q}$, then $[c, {\cal Q}] = c$
requires that ${\cal Q} g = 0$ and $ h = 0$. Since $h=h({\bar z})$ only, $h$
must be a constant, which we already know is unity.

As ${\cal Q}g=0$, $g=g(\vert z\vert)$ only. Then from $[c,c^\dagger]$,
\begin{equation}
1= {1\over 2}(g+\bar g) + {1\over 2} r{d\>\>\over dr}\left({g+\bar
g\over 2}\right)\>,
\label{e4.4}
\end{equation}
where $r=\vert z\vert$.
Since all other commutation relations follow from these two, no additional
equations for $g$ are found. In particular, notice that there is no
constraint on the imaginary part of $g$. For convenience, we shall
choose $g$ to be real, and shall remark on other choices later. Then
\begin{equation}
g = 1-{a^2\over 2r^2}\>,
\label{e4.5}
\end{equation}
where $a$ is some constant yet to be determined. Notice that when $a=0$,
$g=1$ and we are back to eq.~$(\ref{e3.10})$.

Let $\phi_0$ be the ground state. Then by definition $c\phi_0=0$, and
\begin{equation}
z{\partial\phi_0\over \partial z} + g\phi_0 = 0\>.
\label{e4.6}
\end{equation}
Because the ground state is rotationally invariant, solving
eq.~$(\ref{e4.6})$ gives
\begin{equation}
\phi_0 = {a\over \sqrt\pi}{e^{-a^2/(2r^2)}\over r^2}\>,
\label{e4.8}
\end{equation}
with the correct normalization. Notice that when $g=1$, $\phi_0$ has
the singularity at $r=0$ we mentioned at the beginning of this section.
Other eigenstates $\phi_n$ of ${\cal Q}$ can be found by successive
application of $c^\dagger$ on $\phi_0$. For the first few $n$,
\begin{eqnarray}
\phi_1 &=& {\sqrt 2}\left({z\over{\bar z}}\right)g\phi_0\>,
\nonumber \\
\phi_2 &=&
	{1\over\sqrt3}\left({z\over{\bar z}}\right)^2(2g^2+2g-1)\phi_0,
\nonumber \\
\phi_3 &=&
	{2\over3}\left({z\over{\bar z}}\right)^3(g^3+3g^2-1)\phi_0.
\label{e4.9}
\end{eqnarray}
where we have normalized all these states to unity. We therefore
expect that in general
\begin{equation}
\phi_n = \left({z\over{\bar z}}\right)^n G_n(g)\phi_0\>,
\label{e4.10}
\end{equation}
where $G_n(g)$ is a polynomial in $g$. Using eq.~$(\ref{e4.3})$, we see that
it satisfies the recursion relation:
\begin{eqnarray}
G_n(g) &=&
	-{1\over\sqrt{2}}{e^{-2g}\over(1-g)^n}{d\>\>\over
	dg}\left[e^{2g}(1-g)^{n+1}G_{n-1}\right]\>,
\nonumber \\
	&=&
	\left({-1\over\sqrt{2}}\right)^n
		{e^{-2g}\over(1-g)^n}\left\{{d\>\>\over
	dg}(1-g)^2\right\}^n\>e^{2g}\>,
\label{e4.11}
\end{eqnarray}
given $G_0=1$. Notice also that $\phi_n$ has precisely the form we
expected from the heuristic arguments given in $\bf \S 3$.

$\phi_n$ are functions which are defined throughout $\bf
C$. In particular:
\begin{equation}
\lim_{\vert z\vert\to0} \vert\phi_n(z,{\bar z})\vert^2 = 0
\qquad, \qquad
\lim_{\vert z\vert\to\infty}\vert\phi_n(z,{\bar z})\vert^2 = 0\>,
\label{e4.12}
\end{equation}
with the presence of $g$ in eq.~$(\ref{e4.3})$ serving as a short distance
regulator. We no longer have any singularity problems and it is
trivial to show that not only are the $c$ and $c^\dagger$
given in eq.~$(\ref{e4.3})$ adjoints of one another, but
also that the representation of ${\cal Q}$ given in eq.~$(\ref{e3.8})$ truly
corresponds to an hermitian operator. We have therefore found a
faithful representation of the algebra and the formal analysis done
in $\bf \S 3$ is now justified.

In retrospect, the problem that we had with eq.~$(\ref{e3.10})$
could have been anticipated. Using a coordinate represetation of the
algebra, eigenstates of ${\cal Q}$ are proportional to $e^{2i\theta}$,
where $\theta$ is the phase of $z$. Yet, this phase is not well defined at
$r=0$, which is precisely where the singularity of the wavefunctions
constructed from eq.~$(\ref{e3.10})$ occurs. This singularity
can be seen explicitly by setting $g=1$ in eq.~$(\ref{e4.8})$. A
cutoff must be introduce at small $r$, which is essentially the role
that $a$ plays. Also, notice that eq.~$(\ref{e3.10})$, unlike the
creation and annihilation operators for the SHO, does not contain
within themselves a length scale. Yet the wavefunction must have
units of inverse length. A length scale must be therefore be
introduced, which is a role that $a$ also serves.

With the correct representation of the the algebra known, and the
wavefunction of the single vortex states constructed, we can now
return to our premise that the algebra eq.~$(\ref{e3.11})$ is the correct
quantum mechanical description of a quantum vortices in a two
dimensional superfluid. Specifically, the question is whether or not
the algebra eq.~$(\ref{e3.11})$, and the wavefunctions construced from it,
actually does describe the single vortex state in the superfluid. As
we have seen, $\cal Q$ already has all the agreeable properties we
would expect from a vorticity operator. Moreover, the wavefunctions
have precisely the form we expected from heuristic arguments. The
question is whether or not $H_c$ does. One way to answer this question
is to compare the features $H_c$ with some of the properties of
quantum vortices known from classical fluid dynamics. By doing so  we
shall also be able to identify physically the two constants in the
theory: $\epsilon$ and $a$. We begin with the eigenstates of $H_c$.

There are two features of the classical hamiltonian
eq.~$(\ref{e2.6})$ which are essential in deriving the
renormalization group equations for the KT phase transition. First,
the classical energies are proportional to $q^2$, the vorticity of the
vortex. Second, because of the logarithmic interaction, under scaling
$z\to\chi z$, $K\to K - 2e_{kt}\sum^N_{\alpha\ne\beta} q_\alpha q_\beta
\log\vert\chi/a_s\vert$ and gets shifted by a constant. Since the
hamiltonian is defined only up to a constant anyway, it is {\it
effectively\/} invariant under scaling.

For $H_c$ to be the correct vortex hamitonian, it must have
at least these two features in common with the classical hamiltonian.
{}From eq.~$(\ref{e3.15})$ we see that the eigenvalues of $H_c$ is
proportional to $n(n+1)$. As expected, for large $n$ any differences
between this dependence and $n^2$ is negligible. As for the scaling,
let us write $H_c$ in terms of differential operators,
\begin{eqnarray}
H_c &\equiv& \epsilon c^\dagger c\>,
\nonumber \\
	&=&
	-{\epsilon\over 2}
	\left\{
		\vert z\vert^2{\partial^2\>\>\over\partial z\partial {\bar z}}
		+ g{\bar z}{\partial\>\>\over\partial {\bar z}}
		+ (1-g)z{\partial\>\>\over\partial z}
		+ (1-g^2)\>,
	\right\}
\label{e4.13}
\end{eqnarray}
and consider the limit where $r\gg a$. In this limit, $g\to1$ and we
can see that $H_c$ is also invariant under scaling $z\to\chi z$. But only
in this limit. This is to be expected, however. The classical
hamiltonian has a logarithmic singularity when $z^\alpha\to z^\beta$,
and for it to be well defined, the interaction must be regularized. Any
regularization scheme will ruin the scaling behavior at small $r$,
but at large $r$ the classical hamiltonian is approximatedly scale
invariant, which is precisely the properties that $H_c$ has.

As $H_c$, and its eigenvalues, have many of the same properties as the
classical hamiltonian, we were justified in identifying it as the
single vortex hamiltonian. The unknown energy scale $\epsilon$ can
now be identified as $\epsilon_{kt}$. Each of the eigenstates of
$H_c$ thus correspond to the total energy of a single vortex in the
fluid with vorticity $n$. Classically, we would call this the
``self-energy'' of the vortex, as this is the amount of energy needed
to creat a vortex in the fluid. Unlike the
classical self-energy, which is infinite, all eignvalues
of $H_c$ are {\it finite}, however.

Mathematically, $a$ in eq.~$(\ref{e4.8})$ serves as an short distance
regulator. With its presence the wavefunction vanishes at $r=0$ and
there are no problems with the definition of $\theta$. It is,
however, at this point questionable as to whether or not $a$ has any
other physical relevance since its value does not affect the
eigenvalues of $H_c$. To gain some physical insight into what other
role it may play, consider the usual quantum mechanical hamiltonian
\begin{equation}
H_L = -{\hbar^2\over 2m}\nabla^2 = -{2\hbar^2\over
m}{\partial^2\over\partial z\partial{\bar z}}\>.
\label{e4.14}
\end{equation}
The actual hamiltonian for the vortices is $H_c$,
of course, not $H_L$. We consider $H_L$ only to make a connection
with the usual calculation of the average energy of a
vortex. In doing so we find that
\begin{equation}
\langle\phi_0\vert H_L\phi_0\rangle = {\hbar^2\over ma^2}\>,
\label{e4.15}
\end{equation}
and, in terms of $H_L$, we would say that $\phi_0$ is not a state with
zero energy. Correcting this would only involve shifting $H_L$ by a
constant, however. Of more interest is
\begin{eqnarray}
\langle\phi_1\vert H_L\phi_1\rangle &=&6{\hbar^2\over ma^2}\>,
\nonumber \\
\langle\phi_2\vert H_L\phi_2\rangle &=&26{\hbar^2\over ma^2}\>,
\label{e4.16}
\end{eqnarray}
and we find that $\langle\phi_2\vert H_L\phi_2\rangle\approx
(2)^2\langle\phi_1\vert H_L\phi_1\rangle$. Once again the energies
have the correct dependence on charge. Comparing eq.~$(\ref{e4.16})$
with $K$, we therefore identify $a^2 =1/\rho_0$.

Let us now try to calculate $\Delta x\equiv\sqrt{\langle \vert z\vert
^2\rangle -\vert \langle z\rangle\vert^2}$, the rms ``position'' of
the vortex. By symmetry, $\langle z\rangle =0$. However,
\begin{equation}
\langle \vert z\vert^2\rangle \equiv \int_{\bf C} \vert z\vert^2
\vert\phi\vert^2 d^2x \sim a^2\log{R/a}\>,
\label{e4.17}
\end{equation}
where $R$ is the large distance cutoff and represents the
``size'' of the system. Consequently, $\langle \vert z\vert^2\rangle$
diverges logarithmically and we {\it cannot\/} say that the vortex is
localized at any one point in the fluid.

The parameter $a=1/\sqrt\rho_0$ is offen refered to in the literature
as the ``core size'' of a vortex. Although we see that the classical
notion that the vortex is localized at any specific point in the fluid is
incorrect, let us see if we can still nevertheless consider $a$ as an
effective size of the vortex. Consider the function
$\vert\phi_n\vert^2$, which is rotationally symmetric. From
eq.~$(\ref{e4.12})$ we know that all $\vert\phi_n\vert^2$ have a zero at
$r=0$. Although $\vert\phi_0\vert^2$ has no other zeros, other
$\vert\phi_n\vert^2$ will. In fact, $\vert\phi_1\vert^2$ has an
additional zero at $r =a/\sqrt2$; $\vert\phi_2\vert^2$ has two other
zeros at $r =a/(3\pm\sqrt3)$; while $\vert\phi_3\vert^2$ has three
aditional zeros at $r = a/(2.78)$, $r =a/(1.82)$ and $r =a/(0.967)$.
Since $\vert\phi_n\vert^2$ cannot be negative, there are
corresponding local maxima of $\vert\phi_n\vert^2$ at points in
between two zeros. Due to the $1/r^4$ behavior of
$\vert\phi_n\vert^2$, we would expect the largest maxima to be
between $r =0$ and the next nearest zero of the function. For $n=1,
2, 3$, they all fall within $r = a/\sqrt2$. Because the probablity
density is highest for $r \le a/\sqrt2$, we may identify $a/\sqrt2$
as the effective ``core size'' of a vortex. Indeed, with this
identification the independence of the eigenvalues of $H_c$ on its
value can be understood. It is once again a reflection of the
classical hamiltonian. If we scale $a_s\to la_s$, the classical
hamiltonian $K$ gets shifted only by an irrelevant constant. The
classical hamiltonian, like $H_c$, is also effectively independent of
the ``size'' of the vortex core.

The interpretation of $a$ as a core size runs into problems, however,
if we now reconsider the arbitrariness in the choice of $g$. Remember
that we can add to $g$ a function $g_I$ of $\vert z\vert$
which is purely imaginary, $\overline{g_I} = - g_I$, and $c$ and
$c^\dagger$ will still be faithful representations of the vortex
algebra. From eq.~$(\ref{e4.10})$ we see that doing so will alter the ground
state wavefunction $\phi_0$ by the multiplication of a complex phase
$e^{i\xi}$ where
\begin{equation}
g_I = -{i\over 2} r{d\xi\over dr}\>,
\label{e4.18}
\end{equation}
and $g_I$ is pure imaginary.
Therefore, a shift in $g$ is equivalent to a {\it local\/}
gauge transformation of $\phi_0$. Choosing $g$ to be real
corresponds to a choice of guage. This clearly will not change the
convergence properties of $\phi_0$, or any of the other $\phi_n$, as
long as $\vert g_I\vert$ does not diverge faster than $e^{-a^2/r^2}$.
$\phi_n$ will still, of course, be eigenstates of $H$ and ${\cal Q} $
with the {\it same\/} eigenvalues, since they do not depend on $g$.
$\vert\phi_n\vert^2$ does, however, and the probability density
changes with different $g$. Notice also that changing $g$ will result
in changing $a$, what we have identify as the ``core size''.
Consequently, the concept of a definite size of the vortex is
therefore not well defined as it depends on the choice of $g$. We
shall remark further on this later in $\bf\S 7$.

\noindent{\bf \S 7. Conclusion}

By using the observation that we already know which states of the
fluid are of importance, it was straightforward to develope a
quantum mechanical description of vortices in the fluid. What
we ultimately obtained, however, is a single vortex hamiltonian which
is very much different than what one normally encounters. As the
classical equations of vortex motion, Kirchoff's equations, are
themselves drastically different from the usual equations of motion
of Newtonian mechanics, this should not be too surprizing. What is of
more importance and relevance is that $H_c$, and its eigenvalues,
has many of the same properties as the classical hamiltonian. Namely,
that $H_c$ is scale invariant at large $r$ while its eigenvalues are
proportional to the square of the vortex charge, and is independent
of the ``core size'' of the vortex. For these reasons we believe that
$H_c$, and its eigenstates, form an accurate quantum mechanical
description of the single vortex system.

The question then becomes how such an hamiltonian can come about from
the usual microscopic $^4$He hamiltonian. As ours is essentially a
phenomenological description of the single vortex system, this cannot
be answered within this formalism. $H_c$ must instead be seen as an
effective hamiltonian of the fluid, albeit one which differs
drastically from the ones we are accustomed to using. As for how it
can araise, we can only note that at very low temperatures $^4$He is
strongly interacting. $H_c$ may thus araise as an effective hamiltonian of
this strongly interacting theory, especially since vortices
are the result of a collective property of the system. Quantum
mechanically, they represent a definite excitation
state of the quantum fluid (what is sometimes refered to as
a psuedo-particle), while classically they are a specific
configuration of the current flow in an ideal fluid.

We have been calling $\phi_n$ the single vortex wavefunctions. This
term is a somewhat vague, however, and was mainly used as a matter of
convenience. Physically, we should instead interpret each $\phi_n$
as representing an excitation of the superfluid which give rise to a
single vortex with vorticity $n$ in the fluid. Its norm,
$\vert\phi_n\vert^2$, is therefore the number density for the
$^4$He atoms in this state. (The integral of $\vert\phi_n\vert^2$
is then the total number $N_v$ of $^4$He atoms in this state.
Although we have normalized $\phi_n$ to unity, we could have just as well
normalized it to $N_v$.) It would therefore seem that from the
results of $\bf\S 4$ we have uniquely determined the density of the fluid
needed to creat a quantum vortex with vorticity $n$ in the fluid.
Notice, however, that $\phi_n$ is not unique. It is dependent on $g$,
which itself is defined only up to an
arbitrary function of $\vert z\vert$. Different choices of $g$ will
result in different $\phi_n$, although $\vert\phi_0\vert^2 $ is
always uniquely defined. Consequently, as the eigenvalues of $H_c$
and ${\cal Q}$ do not depend upon $g$, there are {\it many\/} different
excitations of the fluid which will give raise to the same vorticity
state. The condition that the superfluid is in a state of definite
vorticity is {\it not\/} sufficient to determine the state of the
fluid uniquely.

Notice that in general $\vert\phi_n\vert$ is not a constant thoughout
the fluid. In the usual theory of vorticies in the superfluid, however, the
density of the fluid was necessarily a constant almost everywhere.
(Only when one gets close to the vortex core does the density
vary.) And it was only in this manner that the analogy
between the superfluid and the classical ideal fluid could be used. We
now see that this condition is unnecessarily restrictive. Due to the
freedom in choosing $g$, there are {\it infinitely\/} many different
excitations of the fluid, not only the one with constant density,
which have the same vorticity and, more importantly, the {\it same\/}
energy.

In this paper we have only consider the behavior of a single vortex
in the fluid. The next step is to develope a many vortex formulation
of this system $\cite{19}$. Since we have annihilation and creation
operators for these vortices, this is straightforwardly done in
complete analogy to the case of the SHO. Once this is accomplished the
statistical mechanics of the many vortex system can be studied within
an algebraic formalism and compared with the results obtained using
the KT analysis of the classical vortex gas.

\begin{center}
{\bf Acknowledgements}
\end{center}

This work was supported in part by the R.O.C. NSC Grant Nos.
NSC84-2112-M-001-022.

\end{document}